\documentclass[preprint, 5p, number, sort&compress, lefttitle]{elsarticle}

\usepackage{amsmath}
\usepackage{amssymb}
\usepackage{graphicx}
\usepackage{bm}
\usepackage{hyperref}
\usepackage{float} 
\hypersetup{
    colorlinks=true,
    linkcolor=blue,
    filecolor=blue,
    urlcolor=blue,
    citecolor=blue,
}

\def\rm{\mathrm}

\def\gev{\text{GeV}}

\def\elv{E_{\mathrm{LV}}}

\def\fund{National Natural Science Foundation of China~(Grant No.~12075003)}

\def\etal{{\it et al.\/}}

\def\declaration{The authors declare that they have no known competing financial interests or personal relationships that could have appeared to influence the work reported in this paper.}

\newcommand{\eq}[1]{Eq.~(\ref{#1})}
\newcommand{\fig}[1]{Fig.~(\ref{#1})}

\usepackage{ulem}
\usepackage{xcolor}
\newcommand{\new}[1]{{\color{black}{#1}}}

\journal{\href{http://arxiv.org/abs/2210.11376}{arXiv:2210.11376}; Published in \href{https://doi.org/10.1088/1361-6471/accebb}{J.~Phys.~G: Nucl.~Part.~Phys. 50 (2023) 06LT01}}

\begin{document}


\begin{frontmatter}

\title{Light speed variation from GRB 221009A}

\author[aff1]{Jie Zhu}
\ead{jiezhu@pku.edu.cn}

\author[aff1,aff2,aff3]{Bo-Qiang Ma\corref{cor1}}
\ead{mabq@pku.edu.cn}
\cortext[cor1]{Corresponding author}

\affiliation[aff1]{
    organization = {School of Physics,
    Peking University},
    city = {Beijing 100871},
    country = {China}}

\affiliation[aff2]{
    organization = {Center for High Energy Physics, Peking University},
    city = {Beijing 100871},
    country = {China}}

\affiliation[aff3]{
    organization = {Collaborative Innovation Center of Quantum Matter},
    city= {Beijing},
    country = {China}}

\begin{abstract}
It is postulated in Einstein's relativity that the speed of light in vacuum is a constant for all observers.
However, the effect of quantum gravity could bring an energy dependence of light speed, and a series of studies on high-energy photon events from gamma-ray bursts (GRBs) and active galactic nuclei (AGNs)
suggest a light speed variation $v(E)=c\left(1-E / E_{\mathrm{LV}}\right)$ with $E_{\mathrm{LV}}=3.6 \times 10^{17} ~\mathrm{GeV}$  \new{or a bound 
$E_{\mathrm{LV}} \ge 3.6 \times 10^{17} ~\gev$}. From the newly observed gamma-ray burst GRB 221009A, we find that a 99.3~GeV photon detected by Fermi-LAT is coincident with the sharp spike in the light curves detected by both Fermi-GBM and HEBS under the above scenario of light speed variation, suggesting an option that this high-energy photon was emitted at the same time 
as a sharp spike of low-energy photon emission at the GRB source. 
Thus this highest energy photon event detected by Fermi-LAT during the prompt emission of gamma-ray bursts 
might be considered as an optional signal for
the linear form modification of light speed in cosmological space.

\end{abstract}


\end{frontmatter}


On 9 October 2022, the Gamma-ray Burst Monitor (GBM)~\cite{GBMintro}
onboard the Fermi Gamma-ray Space Telescope (FGST) 
triggered a special gamma-ray burst (GRB), numbered as GRB 221009A, at $T_0=13:16:59.000 \text{ UT}$~\cite{gcn32636}. 
The prompt emission of this extraordinarily bright GRB 221009A was  detected 
not only by Fermi~\cite{gcn32636,gcn32637,gcn32642,gcn32658,gcn32760}, 
but also by other space observatories such as Swift~\cite{gcn32632,gcn32688}, 
AGILE~\cite{gcn32650,gcn32657}, 
INTEGRAL~\cite{gcn32660}, 
Solar Orbiter ~\cite{gcn32661},
SRG~\cite{gcn32663}, 
Konus~\cite{gcn32668},
GRBAlpha~\cite{gcn32685}, 
STPSat-6~\cite{gcn32746}, 
HEBS~\cite{hebs}, 
and by ground observatories such as LHAASO~\cite{gcn32677} with striking very-high-energy features~\cite{Li:2022wxc}. 
This burst, located at around RA = 288.282 and Dec = 19.495~\cite{gcn32658}, is a long burst but with a very small redshift \(z=0.1505\)~\cite{Redshift1,Redshift} compared to most other long bursts.
It is estimated that there is only a 10\% probability to observe an event like GRB 221009A about 50 years after the discovery of the first GRB~\cite{gcn32793}.

The Fermi Large Area Telescope (LAT)~\cite{LATintro} also detected this GRB.   
The highest energy photon observed by LAT reaches 99.3~GeV (with a probability of 99.2\%)~\cite{gcn32658}, and this photon event represents the highest GRB photon energy ever detected by Fermi-LAT during the prompt emission phase of gamma-ray bursts. We show in this Letter that such a high-energy photon event of Fermi-LAT offers a precious opportunity to explore the light speed variation related with Lorentz violation (LV)~\cite{HeMa}.

It is speculated from quantum gravity that the Lorentz invariance might be broken at the Planck scale ($E_\mathrm{Pl}\simeq 1.22\times10^{19}~\mathrm{GeV}$),
and that the light speed may have a variation with the energy of the photon. 
Amelino-Camelia \etal~\cite{method1,method2} first suggested testing Lorentz violation by comparing the arrival times between high-energy and low-energy photons from GRBs.
For energy $E\ll E_{\rm{Pl}}$, the modified dispersion
relation of the photon can be expressed in a model-independent way as
\begin{equation}\label{eq:mdr}
  E^2=p^2 c^2 \left[1-s_n\left(\frac{pc}{E_{\mathrm{LV}, n}}\right)^n\right],
\end{equation}
from which we have the following speed relation
\begin{equation}\label{eq:speed}
  v(E)=c\left[1-s_n\frac{n+1}{2}\left(\frac{pc}{E_{\mathrm{LV},n}}\right)^n\right],
\end{equation}
where $n=1$ or $n=2$ as usually assumed, $s_n=\pm1$ indicates whether high-energy photons travel faster ($s_n=-1$)
or slower ($s_n=+1$) than low-energy photons, and $E_{\rm{LV},n}$ represents the nth-order Lorentz violation scale.
Jacob and Piran~\cite{formula} suggested a formula of arrival-time difference between astroparticles with different energies as 
\begin{equation}\label{eq:dt}
  \Delta t_{\mathrm{LV}}=s_n\frac{1+n}{2H_0}\frac{E^n_{\mathrm{h}}-E^n_{\mathrm{l}}}{E^n_{\mathrm{LV,}n}}\int_0^z\frac{(1+z')^n\mathrm{d}z'}
  {\sqrt{\Omega_{\mathrm{m}}(1+z')^3+\Omega_{\Lambda}}},
\end{equation}
where $E_{\rm{h}}$ and $E_{\rm{l}}$ correspond to the energies of the observed high-energy and
low-energy photons, $z$ is the redshift of the source GRB, $H_0$, $\Omega_{\rm{m}}$ and $\Omega_{\rm{\Lambda}}$
are cosmological constants. 
\eq{eq:dt} was derived in the standard model of cosmology, and later was found the same as the time-difference formula in expanding Finsler spacetime~\cite{Finsler}.
In this work, we adopt the present day Hubble constant $H_0=67.3\pm 1.2~ \rm{km ~s}^{-1}\rm{Mpc}^{-1}$~\cite{pgb},
the pressureless matter density $\Omega_{\rm{m}}=\mathrm{0.315^{+0.016}_{-0.017}}$~\cite{pgb} and the dark energy density
$\Omega_{\Lambda}=\mathrm{0.685^{+0.017}_{-0.016}}$~\cite{pgb}.
Earlier researches on cosmic photon data of observed GRBs
set robust bounds on Lorentz violation~\cite{Ellis1999sd,Ellis_app,Abdo_2,Xiao:2009xe}.
A series of researches~\cite{Shao2010f,Zhang2015,Xu2016a,Xu2016,
Amelino-Camelia:2016ohi,Amelino-Camelia:2017zva,
Xu2018,Liu2018,Zhu2021a,Chen2021}
on time flights of high-energy photon events from GRBs detected by Fermi Gamma-ray Space Telescope reveal a regularity of high-energy photon events, leading to a light speed variation with $s=+1$, $n=1$, and $\elv=3.6 \times 10^{17} ~\gev$. 
This result suggests a scenario that high-energy photons travel slower than low-energy photons. It is also shown that 
several phenomena related to the light curves of three active galactic nuclei (AGNs), namely Markarian 421 (Mrk 421), Markarian 501 (Mrk 501) and PKS 2155-304, can serve as the supports for the light speed variation determined from GRBs
at a same scale of $\elv=3.6 \times 10^{17} ~\gev$~\cite{Li2020}.
Such a value is compatible with constraints from analyses on variety GRB data~\cite{Bolmont:2006kk,Chang:2015qpa,Bernardini:2017tzu,Pan:2020zbl}, including recent
analyses using eight Fermi-LAT  gamma-ray burst data~\cite{Ellis:2019} and using spectral lag data from the BATSE satellite~\cite{BATSE21}.
It is also found to be consistent with predictions in a string theory model for space-time foam~\cite{Ellis:2008gg,Li:2009tt,Li:2021gah,Li:2021eza,Li:2021tcv,Li:2021tcv2}. 
\new{From a conservative viewpoint, we may consider these results as a suggestion for a lower bound of the Lorentz violation scale $\elv \ge 3.6 \times 10^{17} ~\gev$.}
 
\begin{figure*}
    \centering
    \includegraphics[scale=0.65]{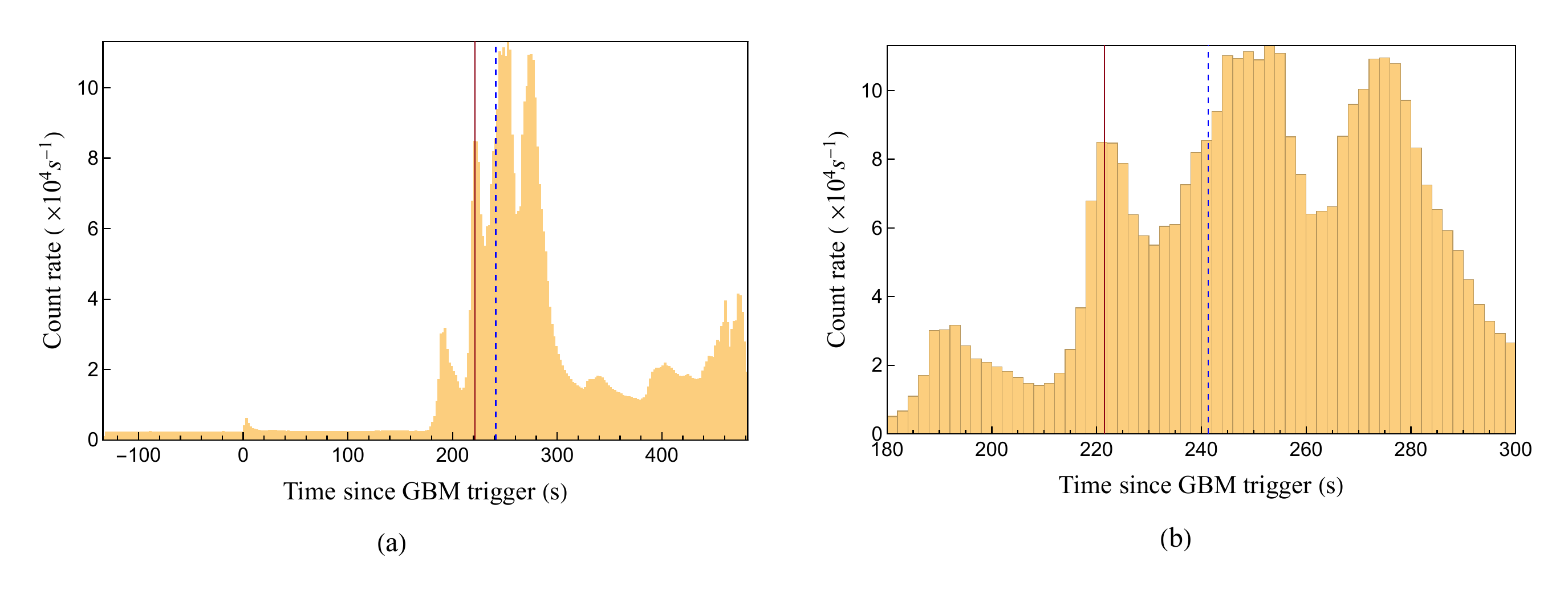}
    \caption{Light curves of Fermi-GBM with the two brightest trigger detectors combined (GBM NaI-n7 and NaI-n8, 7keV-1MeV) for GRB 221009A, with events binned in 2-second intervals. The red vertical line represents $T_1'=221.5~\text{s}$, the estimated position
    of the 99.3~GeV photon event detected by Fermi-LAT if there is no light speed variation, and the blue dashed vertical line represents $T_1=241.3~\text{s}$, the observed position of this photon event.  The red line very well matches a peak of low-energy photons.}
    \label{fig:gbm}
\end{figure*}

As the newly detected GRB 221009A is the brightest one among all detected GRBs and also is the one with a very short distance from the Earth, the light curves of this rare GRB should contain rich information from which we can reveal novel
features of our Universe.
To simplify the discussion, here we treat the GBM trigger time $T_0$ as 0. 
Detailed information of the highest GRB photon energy ever detected by Fermi-LAT during the prompt phase of GRBs can be found at the LAT data server~\cite{lat_data}. The observed energy of this photon is $E=99.3254~\gev$, the position of this photon in J2000 is (287.533, 19.875), and the arrival time is $T_1=T_0+241.326~\text{s}$. Also, the information of low-energy photons can be found at the Fermi server~\cite{gbm_data}, and the light curve of GRB 221009A from two brightest GBM trigger detectors combined (GBM NaI-n7 and NaI-n8, 7keV-1MeV) are shown in \fig{fig:gbm}. 
The Lorentz violation scenario provided above suggests that the low-energy photons (less than 1 MeV) emitted at the same time as the 99.3~GeV photon at the source travel faster than the latter one, and the arrival time difference can be calculated by \eq{eq:dt} as
\begin{equation}\label{eq:td}
    \Delta t=\frac{E}{\elv H_0}\int_0^z\frac{(1+z')\mathrm{d}z'}
  {\sqrt{\Omega_{\mathrm{m}}(1+z')^3+\Omega_{\Lambda}}}= 19.8 \text{ s}.
\end{equation}
Thus the low-energy photons emitted at the same time as the 99.3~GeV photon at the source would be detected at $T_1'=T_1-\Delta t=221.5 ~\text{s}$.
Turn to the light curve of GBM, amazingly we find that at $T_1'=221.5~\text{s}$ (the red vertical line in \fig{fig:gbm}) there is a significant sharp peak of low-energy photons. That may imply that at the source the 99.3~GeV photon was emitted at the same time as the low-energy photons detected by GBM as a sharp spike centered at around $T_1'=221.5~\text{s}$. 
\new{Since the energy resolution of LAT is within 10\% uncertainty and the time resolution for GBM and LAT is smaller than 10~$\mu s$,
the error-bars of the Fermi data can be safely neglected in our analyses. Also, the energy value of the low-energy photons is taken as zero in the time-lag analysis~\eq{eq:td} compared to the 99.3~GeV photon event, so the energy values of the low-energy photons can be neglected. 
}

\begin{figure*}
    \centering
    \includegraphics[scale=0.75]{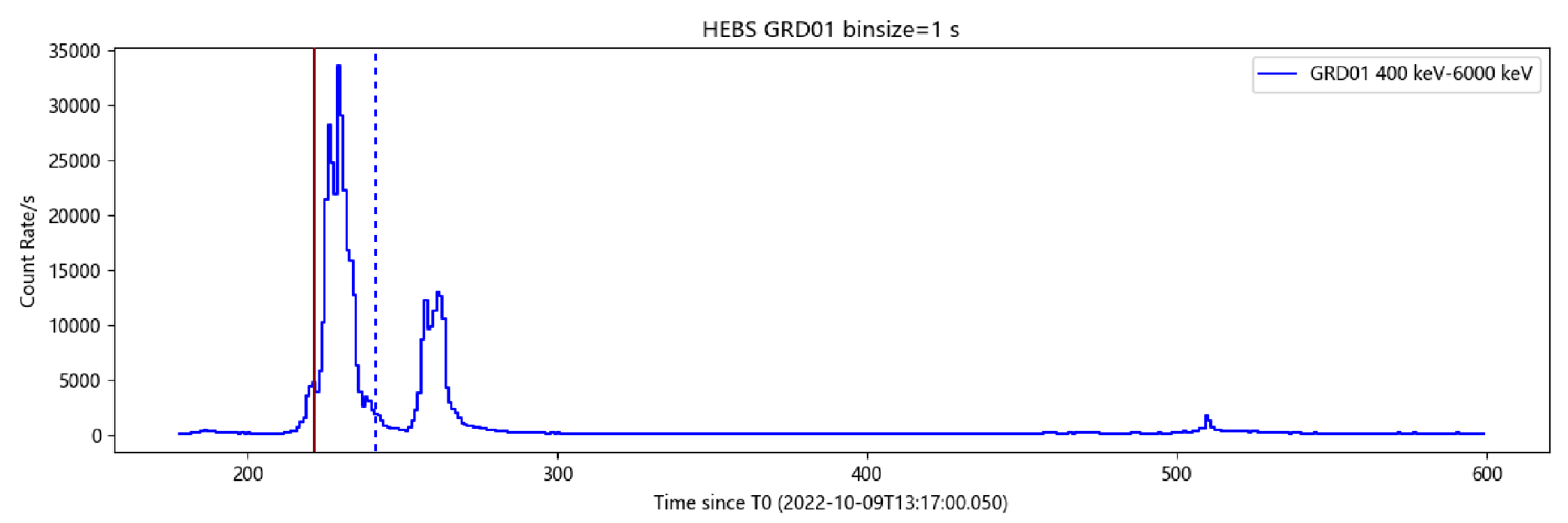}
    \caption{Light curves of HEBS for GRB 221009A from \href{https://gcn.gsfc.nasa.gov/gcn3/32751.gcn3}{GCN Cir.~32751}~\cite{hebs}. The red vertical line represents $T_1'=221.5~\text{s}$, the estimated position
    of the 99.3~GeV photon event detected by Fermi-LAT if there is no light speed variation, and the blue dashed vertical line represents $T_1=241.3~\text{s}$, the observed position of this photon event. The red line also very well matches a peak of low-energy photons.}
    \label{fig:hebs}
\end{figure*}

For the low-energy photon events of GRB 221009A, the HEBS collaboration also published the light curve~\cite{hebs} (\fig{fig:hebs}). Different from GBM, the energy range of HEBS is from 400~keV to 5~MeV. We can also see that there is a tiny sharp peak of low-energy photon events at $T_1'=221.5~\text{s}$ (the red vertical line in \fig{fig:hebs}). Due to the long duration of the GRB, the 99.3~GeV event locates in the emission phase of low-energy photons detected by the Fermi detectors at the Earth, but the actually observed position (the blue dashed line in \fig{fig:hebs}) of this Fermi-LAT event 
is outside the main peak ranges of the HEBS light curves, showing
the difficulty to associate it with the accompanying low-energy photons in the HEBS light curves without considering the Lorenz violation effect.
We can see
that the estimated position  
of this Fermi-LAT event just corresponds to the sharp spike of low-energy photons in both 
GBM and HEBS light curves, as shown with a red vertical line in both \fig{fig:gbm} and \fig{fig:hebs}. The coincidence of this fact 
supports the Lorentz violation scenario of $v(E)=c\left(1-E / E_{\mathrm{LV}}\right)$ with $E_{\mathrm{LV}}=3.6 \times 10^{17} ~\gev$ \new{or a bound $E_{\mathrm{LV}} \ge 3.6 \times 10^{17} ~\gev$}.

In conclusion, we analyze the $99.3$~GeV photon event detected by Fermi-LAT 
at $T_1=241.3$~s after the GRB 221009A trigger time $T_0=0$, and find that in the Lorentz violation scenario of $v(E)=c\left(1-E / E_{\mathrm{LV}}\right)$ with $E_{\mathrm{LV}}=3.6 \times 10^{17}~\gev$, at the GRB source this photon was emitted at the same time as the low-energy photon emission observed as a sharp spike centered at around 221.5~s
in the light curves reported by Fermi-GBM and HEBS. This remarkable  coincidence between the highest energy photon event observed by Fermi-LAT during the prompt phase and the sharp spike in the low-energy light curves of brightest gamma-ray burst GRB 221009A might be considered as an optional signal
for
the light speed variation \new{with $E_{\mathrm{LV}}=3.6 \times 10^{17}~\gev$ or the bound 
$E_{\mathrm{LV}} \ge 3.6 \times 10^{17} ~\gev$}
suggested in previous studies~~\cite{Shao2010f,Zhang2015,Xu2016a,Xu2016,
Amelino-Camelia:2016ohi,Amelino-Camelia:2017zva,
Xu2018,Liu2018,Zhu2021a,Chen2021,Li2020}. \new{Of course, 
more scrutiny is necessary to test the scenario suggested in this Letter when more data are available.}

\section*{Declaration of competing interest\label{declaration}}

\declaration{}

\section*{Data availability statement\label{availability}}
All data that support the findings of this study are referenced within the article (and any
supplementary files).

\section*{Acknowledgements\label{acknowledgements}}

This work is supported by \fund.


\end{document}